\documentclass[preprint]{aastex} 
\usepackage{latexsym}
\usepackage{graphicx}
\usepackage{amssymb}
\usepackage[usenames]{color}
\newcommand{\be}{\begin{equation}}
\newcommand{\ee}{\end{equation}}
\renewcommand{\v}[1]{\mathbf{#1}}

\begin{document}

\title{Eccentricity evolution of giant planet orbits \\
due to circumstellar disk torques}

\medskip
\author{Althea V. Moorhead$^1$ and Fred C. Adams$^{1,2}$} 
\bigskip 

\affil{$^1$Michigan Center for Theoretical Physics \\ 
Physics Department, University of Michigan, Ann Arbor, MI 48109}

\affil{$^2$Astronomy Department, University of Michigan, Ann Arbor, MI 48109}

\begin{abstract}

The extrasolar planets discovered to date possess unexpected orbital elements.  Most orbit their host stars with larger eccentricities and smaller semi-major axes than similarly sized planets in our own solar system do.  It is generally agreed that the interaction between giant planets and circumstellar disks (Type II migration) drives these planets inward to small radii, but the effect of these same disks on orbital eccentricity, $\epsilon$, is controversial.  Several recent analytic calculations suggest that disk-planet interactions can excite eccentricity, while numerical studies generally produce eccentricity damping.  This paper addresses this controversy using a quasi-analytic approach, drawing on several preceding analytic studies.  This work refines the current treatment of eccentricity evolution by removing several approximations from the calculation of disk torques. 
We encounter neither uniform damping nor uniform excitation of orbital eccentricity, but rather a function $d\epsilon /dt$ that varies in both sign and magnitude depending on eccentricity and other solar system properties.  Most significantly, we find that for every combination of disk and planet properties investigated herein, corotation torques produce negative values of $d\epsilon/dt$ for some range in $\epsilon$ within the interval [0.1, 0.5].  If corotation torques are saturated, this region of eccentricity damping disappears, and excitation occurs on a short timescale of less than 0.08 Myr.  Thus, our study does not produce eccentricity excitation on a timescale of a few Myr -- we obtain either eccentricity excitation on a short time scale, or eccentricity damping on a longer time scale.  Finally, we discuss the implications of this result for producing the observed range in extrasolar planet eccentricity.   

\end{abstract}

\medskip 
$\,$  


\bigskip

\section{Introduction}

Our solar system contained the only known planets orbiting a main sequence star until 1995, when a planet was found orbiting the Sun-like star 51 Peg via a periodic Doppler shift in the star's spectrum.  The discovery of more than 150 additional extrasolar planets has overturned our understanding of what constitutes a typical planetary system.  The extrasolar planets discovered to date are Jupiter-sized, possess a wide range of orbital eccentricities ($0 \le \epsilon \le 0.9$), and orbit their host stars with small semi-major axes ($0.03 \mbox{ AU} \le a \le 6 \mbox{ AU}$).   In contrast, similarly sized planets in our solar system (Jupiter and Saturn) live in nearly circular orbits at 5 and 10 AU.

These discoveries prompted corresponding shifts in solar system evolution theory.  We previously believed that planets formed in, or near, their current orbits.  However, ice, which plays a significant role in forming the dense solid core of a gas giant, will not condense within 3 AU of a Sun-like star, which in turn implies that gas giant planets are unlikely to form with the small semi-major axes they possess.  We now believe that massive planets form outside of this ``snow line'' and subsequently move inward via interactions with a circumstellar disk, a process known as planetary migration.  There are two limiting cases in migration theory in a laminar disk: Type I migration, in which a planet lacks sufficient mass to clear a gap in the disk material and is driven inward by a density wake in the disk, and Type II migration, in which massive planets do clear a gap and are driven inward by resonances between the planet and material in the remainder of the disk.  If the disk is not laminar, Type I migration can be overwhelmed by turbulent effects (Laughlin et al. 2004, Nelson 2005).

In the standard description of planetary migration, rings of disk material in resonance with a planet in a nearly circular orbit exert torques on the planet, driving it inward and further decreasing its eccentricity $\epsilon$ (Goldreich and Tremaine 1980, hereafter GT80).  This scenario, while appropriate for our solar system, never allows the planet to possess large orbital eccentricity, and hence does not explain the large eccentricities of many detected planetary orbits.  As a result, one of the current challenges in planetary migration theory is to produce the wide range in orbital eccentricity observed in the extrasolar planet population.  The observed range in eccentricity can be explained by disk torques acting in conjunction with interactions with a second planet (Moorhead and Adams 2005, hereafter MA05; see also Rasio and Ford 1996, Thommes and Lissauer 2003, Adams and Laughlin 2003).  Here we investigate whether eccentricity excitation can take place when we remove the small eccentricity assumption from the GT80 formalism. 

The combination of planet-planet scattering and simplified disk effects (constant damping rates of semi-major axis and eccentricity) successfully reproduces the observed $a$-$\epsilon$ distribution of the extrasolar planets (MA05).  The disk drives both planets inward while interactions between the planets compete with the disk's eccentricity damping; the result is both small $a$ and a large range in $\epsilon$.  In treating the disk as a source of constant eccentricity damping, this past work was based on a host of analytic (e.g., GT80) and numerical (Trilling et al. 1998, Nelson et al. 2000, Papaloizou et al. 2001) studies.  However, several recent studies (Goldreich and Sari 2003, hereafter GS03; Ogilvie and Lubow 2003, hereafter OL03) suggest that disk torques may excite rather than damp eccentricity.  In this paper, we investigate under what conditions nearly Keplerian disks -- like those observed in nearby regions of star formation -- will give rise to eccentricity excitation or damping.  We then outline a method for incorporating the results into numerical simulations in order to investigate to what extent the $a$-$\epsilon$ distribution changes under this modification.

This paper combines the formalism of three key analytic papers on planet migration: GT80, GS03, and OL03.  Specifically, we obtain our torque formulas from GT80, the form of the eccentricity derivative from GS03, and our treatment of saturated corotation resonances from OL03.  We extend this combined treatment primarily by performing a full calculation of the coefficients of the cosine expansion of the planet's perturbation to the overall gravitational potential, $\phi^\mathsf{P}_{\ell ,m}(\beta)$, assuming neither small eccentricity nor large azimuthal wavenumber (as in past studies).  We find that the function $\phi^\mathsf{P}_{\ell ,m}(\beta)$ has a complicated dependence on eccentricity (see section 2).  Each resonant torque depends on $\phi^\mathsf{P}_{\ell ,m}(\beta)$, and as a result, the eccentricity time derivative is also a complicated function of eccentricity; in particular, $d \epsilon / dt$ attains both positive and negative values depending on the current value of eccentricity.  Several properties of the planet's cleared gap, including its width, placement around the planet, and the degree to which it is cleared, alter the shape of $d \epsilon / dt$.  In this paper, we present eccentricity time derivatives for a variety of gap widths and shapes, seeking common behaviors. 
Additionally, if corotation resonances become saturated,  we find $d \epsilon / dt$ to be almost exclusively positive.  In this manner, we attain a greater understanding of how the production of either eccentricity excitation or damping depends on current planet and disk properties.  The paper concludes with a discussion of this finding in the context of recent analytic studies as well as numerical studies, which generally produce eccentricity damping.

\section{Methods and Initial Conditions}

\subsection{Disk Properties}

For the sake of definiteness, we consider a $1 M_\mathsf{J}$ planet orbiting a 1 $M_\odot$ star, embedded in a 0.05 $M_\odot$ circumstellar disk with radius 30 AU.  We assume that the unperturbed disk surface profile density falls off with distance from the central star as a power law, $\Sigma = \Sigma_0 (r/r_0)^{-1/2}$.  We choose this power-law index for comparison with the many numerical studies that do the same; additionally,  the results are largely insensitive to this choice.  We can easily verify that the small chosen disk mass does not disturb the overall Keplerian rotation curve $\Omega(r)$, i.e., the condition
\be
{G \Sigma \over \Omega^2 r} \ll 1
\ee
is satisfied; the largest value $G \Sigma / \Omega^2 r$ attains for any value of $r$ in our given range is 0.04.

We assume that the radial temperature profile also obeys a power law, $T = T_0 (r/r_0)^{-3/4}$, where  $T_0 = 50$ K at the snow-line, $r_0$ = 7 AU.  Finally, we specify the properties of the disk material, the standard viscosity parameter for accretion disks $\alpha = 10^{-3}$ and the ratio of specific heats $\gamma = 1.4$.  These parameters determine the extent to which planets are able to clear gaps in the disk and the strength of the torques produced by disk-planet resonances (see GT80, Shu 1992, Lin and Papaloizou 1993).

We assume a flat (cold) disk in our analysis.  In addition, our torque formulas require that the disk satisfy the more stringent condition,
\begin{equation}
m {h \over r} = m {(c_s / \Omega) \over  r} \ll  1 \, ,
\label{eq:heq}
\end{equation}
for each resonance of order $m$, where $h$ is the disk scale height and $c_s$ is the sound speed.  The largest value $h/r$ attains is 0.07, where $h/r \simeq 0.04 (r/1~\mbox{AU})^{1/8}$.  A $1~M_J$ planet at 1 AU allows for only 5-10 resonances of each type, and so we can safely use the flat disk torque formulas.  Notice, however, that smaller planets produce smaller gaps, which in turn require larger values of $m$.  As a result, this treatment cannot be extrapolated to arbitrarily small planet masses.

\subsection{Disk Torques}

For a Keplerian disk, the eccentricity evolution due to disk torques is given by (GS03)
\be
{d\epsilon \over dt} = {(1-\epsilon^2)\left[{(1-\epsilon^2)^{-1/2}-{\ell / m}}
\right]/\epsilon \over M_P \sqrt{G M_{\displaystyle \ast} a}}T_D \, ,
\ee
where $a$, $\epsilon$, and $M_P$ are the semi-major axis, eccentricity, and mass of the planet, $m$ is the azimuthal wavenumber of the pattern speed (see below), and $T_D$ is the portion of the torque exerted by the 
disk on the planet that corresponds to the Fourier component of the planet's potential with azimuthal wavenumber $m$ and pattern speed $(l/m) \Omega_P$.

The system contains an infinite number of resonances, and corresponding torques, between the disk and the planet.  For example, if we assign these resonances wavenumbers ($m$, $\ell$), the location of a corotation resonance is given by $r_C = a[m/(\ell)]^{3/2}$.  The massive planets of interest here will clear large gaps in the disk, providing, for each value of $\ell$, an upper limit on the number of resonances contributing to the total torque.  The shape of the gap also affects the eccentricity evolution; Lindblad resonances are proportional to surface density, and corotation resonances are proportional to the radial derivative of surface density.  We first calculate the location and torque of each resonance (and discuss the perturbing function's coefficients $\phi^\mathsf{P}_{\ell ,m}(\beta)$ in detail).  We then present our method for translating a given gap profile into an upper limit on the number of contributing resonances.

Resonances occur, and torques are exerted, where the motion of a ring in the disk matches the pattern speed $\Omega_{\ell ,m}$ of the planet.  For a Keplerian disk, this condition takes the form
\be
\Omega_{\ell ,m} = \Omega_P  + (\ell-m) \kappa_P / m = {\ell \over m} \Omega_P \hspace{0.25cm} \, .
\ee
We must consider both Lindblad and corotation resonances in our calculations, as described above.

\subsection{Lindblad Resonances}

Lindblad resonances occur at radii $r$ where $\Omega(r) \pm \kappa(r) / m = \Omega_{\ell ,m}$, where $m>0$.  For a Keplerian disk, this condition can be written in the form
\be
\Omega(r) = \left({m \over m \pm 1}\right) \Omega_{\ell ,m} \, .
\ee
The radius of a Lindblad resonance is denoted $r_L$.  In the above equation and throughout this discussion, we take the top sign for an outer Lindblad resonance and the lower sign for an inner Lindblad resonance.  The expression for the disk torque due to a Lindblad resonance in a cold, Keplerian, non-gravitating disk is given by
\be
T^\mathsf{L}_{\ell ,m} = {m \pi^2 \over 3 (1 \pm m)} \left[{{\Sigma(r) \over \Omega^2(r)} 
\left({{r d \phi^\mathsf{P}_{\ell ,m} \over d r} \mp 2 m \phi^\mathsf{P}_{\ell ,m}}\right)^2}\right]_
{\displaystyle r_L} \, ,
\label{eq:ltorque} 
\ee
where $\phi^\mathsf{P}_{\ell ,m}$ is the ($l$, $m$) component of the cosine expansion of the disturbing potential produced by the planet (see GT80), and is discussed in further detail below.

\subsection{Corotation Resonances}

Corotation resonances occur at radii $r$ where
\be
\Omega(r) = \Omega_{\ell ,m} \, .
\ee
We denote the radius at which we encounter a corotation resonance $r_C$.  The expression for the disk torque due to a corotation resonance (GT80) in a cold, Keplerian, non-gravitating disk is given by
\be
T^\mathsf{C}_{\ell ,m} = -{4 m \pi^2 \over 3} \left[{{r \over \Omega(r)}{d \over d r}\left({
\Sigma \over \Omega}\right) (\phi^\mathsf{P}_{\ell ,m})^2}\right]_{\displaystyle r_C} .
\label{eq:ctorque} 
\ee

\subsection{Determining the Expansion Coefficients}

Just as the disk is not massive enough to be self-gravitating, the orbiting planet is not massive enough to alter the $1/r$ potential produced by the central star.  As a result, we treat the planet's gravitational potential as a small perturbation in the overall potential, and consider the cosine expansion of the potential.  The elements of this expansion correspond to resonances between the planet and disk material at different radii.

The disturbing potential $\phi^\mathsf{P}$ produced by an orbiting planet moving in the plane of the disk is well known (for example, Murray and Dermott 2001), and is given by 
\begin{eqnarray}
\phi^\mathsf{P}(r, \theta, t) & = & G M_P \left({\v{r_P} \cdot \v{r} \over r^3} - {1 \over 
| \v{r} - \v{r_P}|}\right) \\
& = &  G M_P \left({r_P \cos{(\theta - \theta_P)} \over r^2} - {1 \over 
\sqrt{r^2 + r_P^2 - 2 r r_P \cos{(\theta - \theta_P)}}}\right) \, \nonumber
\end{eqnarray}

We can expand in terms corresponding to pattern speeds $\Omega_{\ell ,m}$:
\be
\phi^\mathsf{P}(r, \theta, t) = \sum_{\ell=-\infty}^{\infty}{\sum_{m=0}^{\infty}{\phi^\mathsf{P}_
{\ell ,m}(r) \cos{(m \theta - \ell \Omega_P t)}}} \, .
\ee
Defining $\beta = r/a$, we can write the expansion coefficients $\phi^\mathsf{P}_{\ell ,m}$ in the form
\begin{eqnarray}
\phi^\mathsf{P}_{\ell ,m}(\beta) & = & {G M_P \over 2 \pi^2 a} \int_0^{2 \pi}{\int_0^{2 \pi}
{d\theta }d\xi} \bigg[ \cos{(m \theta - \ell (\xi - \epsilon \sin{\xi}))}(1-
\epsilon \cos{\xi}) \nonumber\\
&& \times \left({{g(\theta,\xi) \over \beta^2}  - {1 \over \sqrt{\beta^2 + (1-
\epsilon \cos{\xi})^2 - 2 \beta g(\theta,\xi)}}}\right) \bigg] \, , \\
g(\theta,\xi) & = & (\cos \xi - \epsilon) \cos{(\theta)}+(\sqrt{1-\epsilon^2} 
\sin{\xi}) \sin{(\theta)} \, . \nonumber
\end{eqnarray}

Note that $\beta$ is not a free variable, but is determined by $m$, $\ell$, and the type of resonance.  As a result, $\phi^\mathsf{P}_{\ell ,m}(\beta)$ is a non-trivial function of $m$, $\ell$, $\epsilon$ and the type of resonance, as both the planet mass $M_P$ and the semi-major axis $a$ are prefactors.

Calculating the coefficients $\phi^\mathsf{P}_{\ell ,m}$ requires the evaluation of an oscillatory two-dimensional integral.  One method of sidestepping this problem is to expand $\phi^\mathsf{P}_{\ell ,m}(r)$ to first order in $\epsilon$ (GT80).  However, extrasolar planets often display large orbital eccentricity, and one goal of migration theory is to explain these large eccentricities.  It is therefore important to consider the validity of the linear approximation of $\phi^\mathsf{P}_{\ell ,m}$ for the full range of eccentricity $\epsilon$.  Figure \ref{fig:compare} displays the behavior of the numerically determined and approximate expressions for one particular coefficient, $\phi^\mathsf{P}_{4,3}(r_{\mathsf{ILR}})$, as a function of eccentricity.  Clearly, an approximation to first order in eccentricity is only useful for eccentricities $\epsilon \lesssim 0.3$.  In the context of GT80, this approximation was acceptable as researchers were mainly interested in explaining the properties of our own solar system, where planets have nearly circular orbits.  However, many extrasolar planets possess large eccentricities, and thus the small $\epsilon$ approximation necessarily breaks down.

An alternative method for avoiding the full calculation of $\phi^\mathsf{P}_{\ell ,m}$ is to take the large $m$ limit, where the Lindblad and corotation resonances have the same functional dependence on $\phi^\mathsf{P}_{\ell ,m}(r)$ (see Eqs. \ref{eq:ltorque} and \ref{eq:ctorque}).  In this limit, one can evaluate the degree to which Lindblad and corotation resonances cancel or add (GS03).  The torque is proportional to $m$, so large $m$ resonances are more significant.  This approximation works well for less massive planets, which clear small gaps and allow large $m$ resonances to contribute to the torque.  On the other hand, a 10 $M_J$ planet clears a wide enough gap that only a few resonances exist, and even a 1 $M_J$ planet only allows for only 5 to 10 resonances of each type.  Thus, an important part of this study is to calculate the coefficients $\phi^\mathsf{P}_{\ell ,m}(r)$ to much higher order accuracy for larger values of eccentricity, $\epsilon$.

To conclude, for the currently observed sample of extrasolar planets, making low $\epsilon$ or large $m$ approximations may not produce the correct eccentricity evolution for any given system; on the other hand, calculating the exact solution for $\phi^\mathsf{P}_{\ell ,m}(r)$ is often not computationally feasible.  Therefore, we embark on a semi-analytic investigation of the eccentricity evolution for intermediate values of eccentricity and for a number of representative disk-planet configurations. We look for an overall pattern, both to understand eccentricity evolution and to incorporate into future calculations of planetary migration.

\subsection{The shape of the function $\phi^\mathsf{P}_{\ell ,m}(\beta)$ and its radial derivative}

Here we note several characteristics of $\phi^\mathsf{P}_{\ell ,m}(\beta)$ and its derivative as a function of $\epsilon$, as shown in Fig. \ref{fig:compare}.  The overall shape consists of three distinct regions: a nearly linear decrease with $\epsilon$ for small $\epsilon$, a cusp located at an intermediate value $\epsilon=\epsilon_0$, and a smooth oscillation for larger values of $\epsilon$.  As $m$ increases, the location of the cusp moves to the left, and the number of oscillations increases roughly in proportion with $m$.  The cusp at $\epsilon_0 = 1- \beta$ is of particular interest because the derivative $d\phi^\mathsf{P}_{\ell ,m}(\beta)/d\beta$ becomes unbounded at $\epsilon = \epsilon_0$.

The function $\phi^\mathsf{P}_{\ell ,m}(\beta)$ is bounded, and we find, as expected (GT80) that the overall amplitude of the resonance falls off as $|\ell-m|$ increases (GT80).
  However, the derivative $d\phi^\mathsf{P}_{\ell ,m}(\beta)/dr$ is unbounded at the location of the cusp in $\phi^\mathsf{P}_{\ell ,m}(\beta)$, $\epsilon_0$.  Furthermore, $\epsilon_0 = |1-\beta| = |1-(m-1)|/\ell$ (for inner Lindblad resonances, for example), and so is dense like the rational numbers.  In other words,
\be
\forall~\epsilon,~h\in\mathbb{R},
\lbrace \phi^\mathsf{P}_{\ell ,m} : ~|\partial_r \phi^\mathsf{P}_{\ell ,m}(\epsilon) \bigg|_{r^\mathsf{P}_{\ell ,m}}|>h \rbrace \neq \varnothing.
\ee

Using this information alone, it is incorrect to ignore higher values of $|l-m|$.  
However, any physical disk will exhibit a smooth response to the planet's stimulus.  To mimic this effect, we manually smooth the potential (and therefore the cusp in $\phi^\mathsf{P}_{\ell ,m}$) over the largest relevant physical scale. 
The viscous length scale is given by
\be
\delta_{\nu} = \left({ \nu \over m [-  d\Omega / dr]}\right)^{1/3} \, ,
\ee
where $\nu$ is the characteristic kinematic viscosity (OL03).  For $m=1$ and $r=1$ AU, $\delta_{\nu} = 0.004$, and for any $m$, $r$ of interest, $\delta_{\nu} \lesssim 0.01$.  On the other hand, the size of the disk scale height is $h/r \simeq 0.04$.  We will present results using first the viscous length scale as our chosen smoothing length, and then using the disk scale height.  This is to demonstrate the effect that increasing the smoothing length has on our results.
   Smoothing over these physical scales prevents the derivative from attaining arbitrarily large values, and higher order resonances may be safely discarded. 

Furthermore, we may apply a low order correction for our treatment of the disk as two dimensional by averaging over the disk scale height (Menou and Goodman, 2004).  The disk scale height is about four times as large as the viscous scale length, and therefore we will discard the viscous scale length and smooth all of our functions over the disk scale height.

It is important to note that the amplitude of the resonances decreases much more rapidly for small $\epsilon$ than for large $\epsilon$.  Thus, it is important to include resonances for which $|\ell - m| > 1$ in order to calculate $d \epsilon / d t$ accurately.  However, the shape of $\phi^\mathsf{P}_{\ell ,m}(\beta)$ does not change rapidly with $|\ell - m|$ (see Fig. \ref{fig:highell}), and higher order resonances tend to add to a similar shape as do lower order resonances.  Therefore, we take the approach of using $|\ell - m| \le 1$ resonances to determine the shape of $d \epsilon / d t$, and then calculating $d \epsilon / d t$ more accurately for key values of $\epsilon$ using all contributing $\ell$, $m$.

\section{Results}

Using the framework developed in the above section, we can calculate the eccentricity time derivative for a given Jovian planet in a given thin disk.  The most important parameters in this calculation are the mass, semi-major axis, and eccentricity of the planet, and the surface density profile in the vicinity of the planet's cleared gap.  Our algorithm can be summarized as follows: [1] We define the radial surface density profile of the disk and the orbital properties of the planet. [2]  We evaluate the formulae for $\phi^\mathsf{P}_{\ell ,m}$ and its derivative (using the Gauss-Kronrod numerical integration function in \emph{Mathematica}), the torques, and the eccentricity derivative resulting from each resonance, smoothing over the disk scale height to partially account for the disk's finite thickness (Menou and Goodman 2004). [3] We sum the contributions to the total eccentricity derivative over all resonance types and values of $m$.  For each set of input parameters (disk properties plus orbital parameters of the planet), we obtain $d\epsilon/dt$ as a function of $\epsilon$.  Issue may be taken with any individual surface density profile we present in this section.  However, we present $d \epsilon / dt$ for a variety of gap architectures, noting shared characteristics.

\subsection{A Sharp-Edged Gap}

We first present the simplest case: a 1 $M_J$ planet orbiting at $a=1$ AU in a gap with sharply defined edges.    The strongest resonances occur for values of $\ell = m$ or $\ell = m \pm 1$, limiting the number of values of $\ell$ we must consider (GT80).  The azimuthal wavenumber $m$ may still take on any value, but, as $m$ increases, the location of the resonance approaches the semi-major axis of the perturbing planet.  When the planet clears a clean gap, the edges of the gap place a physical upper limit on the number of values of $m$ that must be considered.  The resulting function $d \epsilon / dt$ versus $\epsilon$ (shown in Figs. \ref{fig:oexample} and \ref{fig:example}) is the result of summing the contributions of five to ten resonances of each type (where type refers to, for example, an $\ell = m+1$ inner Lindblad resonance).  

In Fig. \ref{fig:oexample}, $d\epsilon / dt$ is not smoothed over the disk scale height, but is instead smoothed over the viscous length scale $\delta_{\nu} \simeq 0.01$.  Due to the small smoothing length, the features of the $\phi^\mathsf{P}_{\ell ,m}$ functions are clearly visible in Fig. \ref{fig:oexample}; the peaks correspond to the singularities in $d \phi^\mathsf{P}_{\ell ,m}(\beta) / d\beta$ smoothed over the viscous length scale, and the smooth right half of the plot is produced by the combination of the smoothly oscillating portions of the $\phi^\mathsf{P}_{\ell ,m}(\beta)$ and $d \phi^\mathsf{P}_{\ell ,m}(\beta) / d\beta$ functions.
In Fig. \ref{fig:example}, the features of the $\phi^\mathsf{P}_{\ell ,m}$ functions are smoothed out over the disk scale height, $h/r \simeq 0.04$, and thus not clearly visible in the resulting function $d \epsilon / dt$.  However, the peaks and troughs in  $d \epsilon / dt$ for $\epsilon < 0.5 $ generally correspond to the singularities in $d \phi^\mathsf{P}_{\ell ,m}(\beta) / d\beta$, and peaks and troughs in $d \epsilon / dt$ for $\epsilon > 0.5 $ correspond to the smoothly oscillating portions of the $\phi^\mathsf{P}_{\ell ,m}(\beta)$ and $d \phi^\mathsf{P}_{\ell ,m}(\beta) / d\beta$ functions.  The most important result obtained from this plot is that planet-disk interactions do not result in uniform eccentricity damping or excitation.  The behavior of $d \epsilon / dt$ is complicated, leading to varying levels of damping or excitation depending on the current value of $\epsilon$.  

There exist sizable bodies of both analytic and numerical work on the subject of planet-induced gaps in circumstellar disks.  In the context of this study, these works translate into a wide array of possible inputs for the surface density in the vicinity of the planet.  A planet on an eccentric orbit is likely to have a wider gap than a planet in a circular orbit, as is assumed in many of these studies, and so we also perform a calculation of $d\epsilon / d t$ for an artificially widened gap.  The shape of the surface density profile has two important effects on the resulting eccentricity evolution equation $d \epsilon /dt$: [1] Since each torque depends on either the surface density or its radial derivative, altering the shape of the surface density profile will (linearly) change the relative heights of the features in the function  $d \epsilon /dt$.   At the same time, if the width of the gap is unchanged, the overall pattern will also be largely unchanged.  [2] If we instead alter the gap width, the shape of the function $d \epsilon / dt$ changes as resonances are included or excluded.  To understand the general implications of gap width for eccentricity evolution, we present gap architectures with varying gap widths.

The result of slightly narrowing the cleared gap -- a change in the sign of $d \epsilon / d t$ -- is also displayed in Fig. \ref{fig:example}.  However, we are most interested in calculating $d \epsilon /d t $ for intermediate values of eccentricity (and this is the region where our assumptions are most accurate), so this change is not significant within the context of our calculations.

\subsection{The Effect of Residue in the Gap}

When some material remains within the gap, the radius at which disk material no longer exerts torques on the planet is unclear.  To investigate eccentricity evolution in such a gap, we choose a gap shape resembling that produced by the numerical calculations of Bate et al. (2003), where residual material remains within the gap (see Fig. \ref{fig:mybate}).  We then raise the limit on the number of contributing resonances (by raising the allowed values of $m$) until $d \epsilon /dt$ converges to a single value.

As mentioned, the radius at which disk material no longer exerts torques on the planet is unclear.  Therefore, we present $d \epsilon /dt$ with and without torques resulting from coorbital disk material.  Figure \ref{fig:nocorb} displays $(d \epsilon /dt )/ \epsilon$ as a function of $\epsilon$ where only non-coorbital resonances are allowed to contribute.  We find that $(d \epsilon /dt) / \epsilon$ does not converge as $\epsilon$ approaches zero, but instead approaches $+\infty$.  When we include the effects of coorbital resonances (Fig. \ref{fig:coorb}), $d \epsilon /dt$ switches sign for small $\epsilon$, resulting in eccentricity damping rather than excitation.  Notice, however, that in both cases, eccentricity damping results for all values of eccentricity in the range $0.1 < \epsilon < 0.8$.

\subsection{Other configurations}

As mentioned previously, the overall shape of the function $d \epsilon / dt$ versus $\epsilon$ is fairly uniform for all surface density profiles that contain a sharp-edged gap of a given width centered on the planet.  However, there are a number of scenarios in which a planet may lie off-center in the gap: accretion may clear the inner disk, photoevaporation or a close encounter with another star may strip the outer disk, or multiple planets may produce overlapping gaps in the disk.  Of these three possibilities, the situation in which two planets have overlapping gaps is of the greatest interest because two planets orbiting in a close resonance are certain to produce a gap in which neither planet lies at the center.  In contrast, it requires more fine-tuning for other processes to deplete the disk from one of its edges exactly to the semi-major axis of the planet.

To study the effect of a two-planet gap, we place two 1 $M_J$ planets in orbits at $a_1 = $1 AU and $a_2 = $1.59 AU, and thus in a 2-1 mean motion resonance.  The resulting gap in disk material is displayed in Fig. \ref{fig:not}, and the eccentricity evolution it produces is displayed in Fig. \ref{fig:2-1}.  We find that the gap asymmetry does not affect the planets equally; while the inner planet experiences a drastic change in both the shape and magnitude of its $d\epsilon / dt$ function, the outer planet's eccentricity evolves in much the same way a lone planet's eccentricity evolves.  We find, upon closer inspection, that this result is due to the fact that the contribution to $d\epsilon/dt$ from the $m=2$, $\ell=1$ corotation and Lindblad resonances are significantly greater than any other resonance contributions for a single planet, and that these resonances are located on the outer side of the gap.  For the inner planet in a 2-1 resonance, there is no material at this location, but for the outer planet, the location of the 2-1 resonance lies within the disk material.  We can deduce from this that, for any planet orbiting in a disk, a particularly strong resonance exists between the planet and a ring of disk material inside the planet's orbit. It is then important to combine this effect with that of the mean motion resonance (Murray and Dermott 2001, Thommes and Lissauer 2003, Lee and Peale 2002) between the two planets when performing semi-analytic studies of solar system evolution such as that in MA05. 

If these results are to be used in future numerical calculations such as MA05 it is important to understand how the eccentricity evolution depends on orbital parameters.  By completing a large range of calculations, we have determined the following: [1] As the semi-major axis $a$ decreases, the surface density in the vicinity of the planet increases, and the planet sits in a deeper, narrower trough.  The function $d \epsilon / dt$ versus $\epsilon$ obtains an increased number of peaks and troughs, smoothed over the disk scale height, in the small $\epsilon$ region as we include more resonances, and an increased magnitude of $d\epsilon / dt$ everywhere due to the larger surface density at small radii.  As $a$ increases rather than decreases, these effects are reversed.  [2] As the planet mass decreases, the planet clears a narrower and narrower gap.  Once again, the function $d \epsilon / dt$ versus $\epsilon$ obtains an increased number of peaks and troughs, again smoothed over the disk scale height, and experiences an decreased overall magnitude due to the dependence $d \epsilon/dt \propto M_P$.  (This dependence arises from Eqs. 3 and 11; In Eq. 3, the formula $d \epsilon/dt$ explicitly contains the term $1/M_P$, and we see from Eq. 11 that ${(\phi^\mathsf{P}_{\ell ,m})}^2$ contributes $M_P^2$.)

\subsection{Saturation of Corotation Resonances}

The torque of the planet on resonances in the disk deposits or removes angular momentum from the resonance locations in the disk.  Lindblad resonances lose this change in angular momentum through a wave flux, but corotation resonances do not.  As a result, they can become saturated and thereby exert a reduced torque on the planet.  In this section, we assume that coorbital corotation resonances will be completely saturated, and thus provide no contribution to the overall eccentricity evolution (Balmforth and Korycansky, 2001). We account for the effects of corotation saturation using the prescription of OL03.

The degree of saturation for a particular ($\ell$, $m$) non-coorbital corotation resonance can be expressed in terms of a single parameter $p$ (OL03), which is given by
\be
p \equiv \left({-d \ln r \over d \ln \Omega}\right) {\phi^\mathsf{P}_{\ell ,m}(\beta) \over \kappa^2} 
\left({ [-  d \Omega / dr] \over \nu / m}\right) \, ,
\ee
where the viscosity $\nu$ is related to the parameter $\alpha$ by the standard Shakura-Sunyaev formulation, $\nu = \alpha \Omega h^2$.  Therefore, assuming the disk is Keplerian, and that the disk scale height $h = c_s / \Omega$, the expression for $p$ becomes
\be
p = {2 \over 3} \phi^\mathsf{P}_{\ell ,m}(\beta) \left[{{3 m_H \over \alpha \gamma k_B T_0}
\sqrt{r \over G M_{\displaystyle \ast}} (r / r_0)^{3/4} }\right]^{2/3} \, .
\ee
Using the quantities specified previously, we obtain $p \approx 0.284 ~ \phi^\mathsf{P}_{\ell ,m}(\beta)$.  Our sample calculation of $\phi^\mathsf{P}_{4,3}(r_{\mathsf{ILR}})$ has an amplitude of about 5 ($\phi^\mathsf{P}_{4,3}(r_{\mathsf{C}})$ has a similar amplitude), and so the saturation parameter $p$ attains values of order unity for some values of $\epsilon$.  Thus, it is important to investigate the  effect that saturation may have on our results.

The torque exerted in the corotation region is reduced by a factor $f(p)$, where $f(p) \approx 0.4019 p^{-3/2}$ for $p \gg 1$, and $f(p) \approx 1-2.044 p^2$ for $p \ll 1$ (OL03).  We interpolate between these two limits in a manner similar to that in GS03 to obtain 
\be
f(p) = (1 + 0.3529 p^3)^{5/6}/(1 + 1.022 p^2)^2 \, .
\ee
The result of including saturation effects is shown in Figs. \ref{fig:nocorbsat} and \ref{fig:scmc}.  Thus, the calculated saturation level of corotation resonances results in  eccentricity excitation over the full range of eccentricity.  If coorbital resonances are allowed to contribute, a small interval of eccentricity damping exists for $\epsilon \lesssim 0.05$.  On a side note, we learn that the substantial eccentricity damping in the low $\epsilon$ limit in Fig. \ref{fig:scmc} is due to coorbital Lindblad resonances and not coorbital corotation resonances.

Taken at face value, our results imply that all single planets starting with $\epsilon > 0.1$ will be ejected from their systems or accreted on time scales shorter than 0.15 Myr.  Furthermore, uniform eccentricity excitation would lead to no multiple planet systems (MA05), in contrast with the observed population of extrasolar planets which contains about 10-20\% multiple systems.  As a result, it is unlikely that corotation resonances are saturated to the degree described above.  The determination of the degree of saturation is thus an important issue for the future.

\subsection{Eccentricity Damping at $\epsilon = 0.3$}

As mentioned earlier in this work, our goal is to compute the eccentricity time derivative for a variety of orbital parameters and gap shapes, looking for a common pattern.  We find that for each set of inputs, the eccentricity time derivative for a single planet without saturation of corotation resonances is negative for some range in eccentricity between 0.1 and 0.5, and is always negative for $\epsilon = 0.3$.  On the other hand, the eccentricity derivative for the same set of parameters, including $\epsilon = 0.3$, is always large and positive when corotation resonances become saturated.

At this point in our study, we narrow our focus to $\epsilon = 0.3$ only.  We now include resonances in our calculation for which $|\ell - m| > 1$ and recalculate $d\epsilon / d t$ for our different gap shapes.  The result is displayed in Table \ref{table:edots}.  The corresponding gaps in surface density are displayed in Fig. \ref{fig:gaps}.  We see that $d\epsilon / d t$ maintains its sign while adopting a greater magnitude in all cases, as expected.

It is clear that, using our approach and assumptions, we obtain, for intermediate eccentricity, either eccentricity damping or extremely rapid eccentricity excitation.  Eccentricity damping at $\epsilon = 0.3$ will prevent planets from attaining the full range of eccentricity values that the observed sample of extrasolar planets possesses.  On the other hand, if corotation saturation occurs according to the formula of Ogilvie and Lubow (2003), planets would attain very large eccentricities in a few thousand years, and thus would likely be ejected from their host systems within the Myr lifetime of the disk.  Either some additional component (such as vorticity of the gap edges) must be included in calculations of eccentricity evolution, or some additional physical phenomenon, such as interactions between planets in multiple planet systems, must take place. 

\section{Summary and Discussion}

In this paper, we
have re-examined the manner in which circumstellar disks exert torques on
giant planets and thereby change their orbital eccentricity.
We have shown that both the small $\epsilon$ and large $m$ approximations for calculating eccentricity evolution are invalid for studying the current population of observed extrasolar planets.  We have removed both of these approximations from the formalism and completed a full calculation of $\phi^\mathsf{P}_{\ell ,m}(\beta)$ (see Fig. \ref{fig:compare}).  The resulting function $d\epsilon /dt$ attains both positive and negative values (see Figs. \ref{fig:oexample} and \ref{fig:example}), reflecting its nature as a composition of $\phi^\mathsf{P}_{\ell, m}(\beta)$ functions.   Thus, Type II migration produces neither uniform eccentricity excitation nor uniform eccentricity damping, but can produce either excitation or damping depending on the combination of disk and planet properties.  We have found that $d\epsilon /dt$ is influenced by gap properties, in particular gap width.  As the gap width changes, the number of included torques changes and we see corresponding addition or subtraction of spikes in the $d\epsilon /dt$ function.  The placement of the planet within the gap is also important.  A planet in a 2-1 resonance with a second planet and thus far offset from the gap center, for instance, may experience substantially different eccentricity evolution than a planet roughly at the center of the gap.

A gap with sharp edges places an upper limit on the number of contributing resonances.  The exact width of such a gap strongly affects the eccentricity time derivative in the low eccentricity limit.  For instance, a small change in the gap width produces eccentricity excitation rather than damping for small eccentricity (Fig. \ref{fig:example}).  Therefore, it is important to understand the shape of the planet's cleared gap to predict the eccentricity evolution in the small eccentricity regime using our approach.  If the gap contains residual contributing material (Figs. \ref{fig:nocorb} and \ref{fig:coorb}) or if the gap is generated by two, rather than one, planets (Fig. \ref{fig:2-1}), the shape of $d\epsilon/dt$ versus $\epsilon$ changes drastically.  

This work has several limitations.  First, we have assumed throughout that the disk is infinitesimally thin.  The accuracy of our method relies on the inclusion of relatively few disk torques.  As $m$ increases, the flat-disk approximation for disk torques is less accurate.  This effect is of concern in our analyses when $\epsilon \lesssim 0.1$.
Second, our calculations converge extremely slowly for large eccentricity.  Furthermore, while D'Angelo et al (2006) showed that the primary effect of moderate planet eccentricity was to widen the gap in the disk, the effects of a very eccentric planet ($\epsilon > 0.5$) on disk geometry are uncertain.
  With these limitations in mind, we present again our $d\epsilon/dt$ plots with data shown only for $0.1 < \epsilon < 0.5$ (Fig. \ref{fig:goodrange}).  This plot demonstrates that, as long as corotation resonances are unsaturated, eccentricity damping necessarily occurs for $0.2 < \epsilon < 0.5$.  Therefore, disk-planet interactions alone cannot produce the full observed range in extrasolar planet eccentricities, assuming corotation resonances are unsaturated.  

If, on the other hand, we find that if corotation resonances are saturated, we obtain almost exclusive eccentricity excitation (Figs. \ref{fig:nocorbsat} and \ref{fig:scmc}).  However, the resulting excitation is so severe that a planet could only remain in orbit about its host star for less than a tenth of a million years (we obtain this timescale when we include all values of $\ell$, $m$ in our $d \epsilon / d t$ calculation).  Assuming that disks last for of order one million years, this scenario fails to explain how planets survive disk-planet interactions.

There are several possible refinements and extensions of this work.  In our treatment of the disk as strictly Keplerian, we neglect the effects of vortices near the gap edges.  A careful treatment of vorticity may alter the strength of the corotation resonances and lead different values of the eccentricity derivative.  Furthermore, it may be informative to extend the range in eccentricity for which this approach is valid, as the observed range in eccentricity for the extrasolar planets ranges from 0 to 0.9.
As mentioned, we have assumed an infinitesimally thin disk throughout this project.  Accounting for the disk's finite thickness is necessary for understanding the behavior of $d \epsilon/dt$ for $\epsilon \lesssim 0.1$. 
Second, as mentioned previously, the function $d \epsilon / dt$ may behave differently than we have calculated here for high eccentricities ($\epsilon \gtrsim 0.5$).  Our calculation will therefore benefit from an improved understanding of disk properties when a massive planet is present in a highly eccentric orbit.  Third, additional disk effects, such as turbulence, may reduce the degree of corotation resonance saturation proposed by Ogilvie and Lubow to a level consistent with eccentricity growth on a timescale of a few million years, allowing planetary orbits to survive Type II migration.
Finally, these results can be utilized in numerical studies of planet migration, such as MA05, where the effects of the disk can be combined with other solar system phenomena, such as secular resonances between planets.

As this set of calculations illustrates, different disk properties, gap properties, and planet properties can lead to different types of behavior concerning the eccentricity evolution of giant planet orbits. In the vast ensemble of star and planet forming environments in our Galaxy (and others), we expect a wide variety of disk surface density profiles, gap widths, gap edge shapes, and other characteristics that determine eccentricity damping and excitation. In addition, for any given set of disk/planet properties, the eccentricity damping and/or excitation rates are complicated functions of eccentricity. As a result, although we have found a common denominator of eccentricity damping  at $\epsilon=0.3$ for a variety of gap shapes and widths, the question of how circumstellar disks affect the eccentricity evolution of their planets should generally be considered only in a statistical sense.

\bigskip   
\centerline{\bf Acknowledgments} 
\medskip 

We would like to thank Ed Thommes for beneficial
discussions and the referees for their many helpful suggestions regarding this paper.  This work was supported at the University of
Michigan by the Michigan Center for Theoretical Physics and by NASA
through the Terrestrial Planet Finder Mission (NNG04G190G), the
Astrophysics Theory Program (NNG04GK56G0), and the Origins of the
Solar System Program.

{}

\begin{table}

\begin{center}
\begin{tabular}{ccccc}
  \parbox{1in}{
    \center
    \linespread{1.0}
    \selectfont
    Gap \linebreak 
    Architecture} & 
  \parbox{1in}{
    \center
    \linespread{1.0}
    \selectfont
    Coorbital \linebreak Material \linebreak Included? \linebreak \smallskip} & 
  \parbox{1in}{
    \center
    \linespread{1.0}
    \selectfont
    Corotation \linebreak Saturation \linebreak Included? \linebreak \smallskip} & 
  $|\ell - m| \le 1$ only & 
  all $\ell$, $m$ \\

\hline
  \parbox{1in}{
    \center
    \linespread{1.0}
    \selectfont
    wide and sharp-edged} 
 & -- & no & -0.16 & -0.05\\
   \parbox{1in}{
    \center
    \linespread{1.0}
    \selectfont
    \textcolor{ForestGreen}{narrow and sharp-edged} }
 & -- & no & -0.17 & -0.048 \\
   \parbox{1in}{
    \center
    \linespread{1.0}
    \selectfont
    \textcolor{blue}{narrowest and sharp-edged} }
     & -- & no & -0.13 & -0.073 \\
   \parbox{1in}{
    \center
    \linespread{1.0}
    \selectfont
    \textcolor{red}{Bate et al. gap} }
     & no & no & -0.052 & -0.025 \\
   \parbox{1in}{
    \center
    \linespread{1.0}
    \selectfont
    \textcolor{red}{Bate et al. gap} }
     & yes & no & -0.050 & -0.019 \\
   \parbox{1in}{
    \center
    \linespread{1.0}
    \selectfont
    \textcolor{ForestGreen}{narrow and sharp-edged} }
     & -- & yes & 0.063 & 0.029 \\
   \parbox{1in}{
    \center
    \linespread{1.0}
    \selectfont
    \textcolor{blue}{narrowest and sharp-edged} }
     & -- & yes & 0.077 & 0.031 \\
   \parbox{1in}{
    \center
    \linespread{1.0}
    \selectfont
    \textcolor{red}{Bate et al. gap} }
     & no & yes & 0.032 & 0.017 \\
   \parbox{1in}{
    \center
    \linespread{1.0}
    \selectfont
    \textcolor{red}{Bate et al. gap} }
     & yes & yes & 0.032 & 0.017 \\
\end{tabular}
\end{center}

\vskip 1.0truein 

\caption{Here we present the timescale for growth or decay in eccentricity resulting from our calculations.  All numbers presented are in units of Myr, and negative values denote eccentricity damping while positive values denote eccentricity excitation.  We present both the results of calculations including resonances with $|\ell - m| \le 1$ and including all contributing resonances.  We find that including resonances for which $|\ell - m| > 1$ decreases the timescale by approximately a factor of 3, but does not change any cases from eccentricity damping to eccentricity excitation or vice versa. 
}
\label{table:edots}

\end{table}

\begin{center}

\begin{figure}
\epsscale{0.6}\plotone{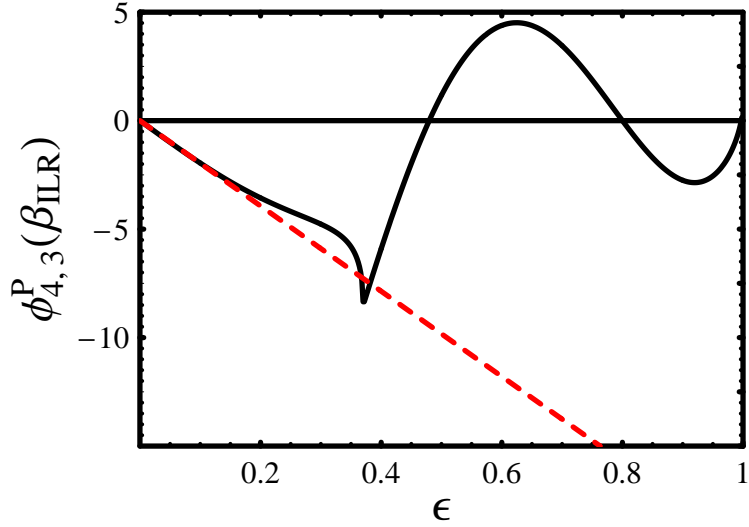}
\caption{The exact solution for $\phi^\mathsf{P}_{4,3}(r_{\mathsf{ILR}})$ (solid 
black curve) and an approximation for $\phi^\mathsf{P}_{4,3}(r_{\mathsf{ILR}})$ 
accurate to first order in eccentricity (dashed red line) as a 
function of eccentricity.  This plot shows that, for this $\ell$, $m$, the linear approximation is 
useful only in the region $\epsilon \lesssim 0.3$.  As the azimuthal wavenumber $m$ increases, the 
cusp in the exact solution moves leftward while the oscillations in 
the right half of $\phi^\mathsf{P}_{\ell ,m}(r_{\mathsf{ILR}})$ increase in number 
and decrease in amplitude, thus shrinking the region of accuracy for 
the linear approximation.}
\label{fig:compare}
\end{figure}

\newpage
\begin{figure}
\plotone{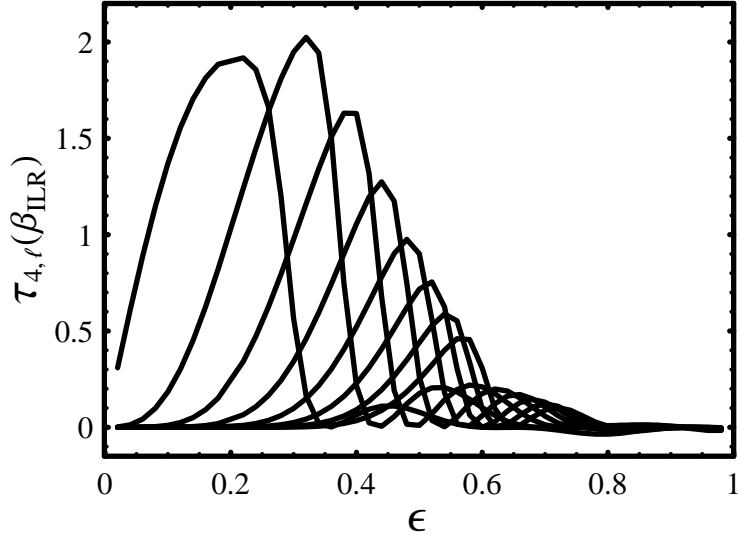}
\caption{The exact solution for the torque exerted by the inner Lindblad resonance $\tau_{\ell, 4}(r_{\mathsf{ILR}})$ as a function of $\epsilon$ for $\ell = $ 5, 6, 7, 8, 9, 10, 11, 12.  As $\ell$
increases, the first peak decreases and moves toward higher eccentricity.  We find that by including resonances for which $|\ell - m| > 1$, we are likely to obtain $d \epsilon / d t$ with increased magnitude, but are unlikely to significantly alter the shape of $d \epsilon / d t$ as a function of $\epsilon$.}
\label{fig:highell}
\end{figure}

\newpage
\begin{figure}
\plotone{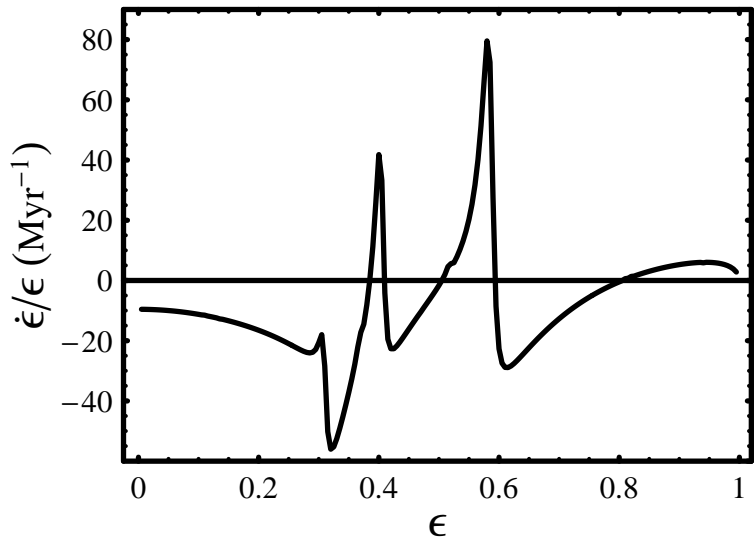}
\caption{The function $d \epsilon / dt$ versus $\epsilon$ for a 1 $M_J$ planet in a 1 AU orbit.   Here we have smoothed over the viscous scale length to demonstrate the shape of $d \epsilon / dt$ for a small smoothing length.  The features of this function clearly result from the $\phi^\mathsf{P}_{\ell ,m}$ functions; each sharp peak and trough occurs at the location of a cusp in  some $\phi^\mathsf{P}_{\ell ,m}$ function, which is the same location at which the radial derivative of $\phi^\mathsf{P}_{\ell ,m}$ becomes unbounded (e.g., see Fig. \protect\ref{fig:compare}).}
\label{fig:oexample}
\end{figure}

\newpage
\begin{figure}
\plotone{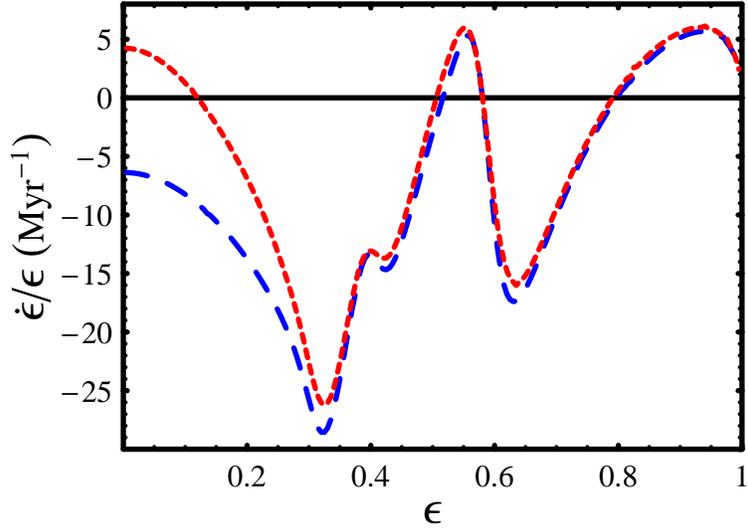}
\caption{The function $d \epsilon / dt$ versus $\epsilon$ for a 1 $M_J$ planet in a 1 AU orbit.  Here we have smoothed over the disk scale height to account for the finite thickness of the disk.  The overall shape of this function is common for all surface density profiles that have gaps with sharp edges centered on the orbiting planet.  The features of this function result from the $\phi^\mathsf{P}_{\ell ,m}$ functions; peaks and troughs for $\epsilon < 0.5$ occur at the locations of smoothed cusps in the $\phi^\mathsf{P}_{\ell ,m}$ functions, and result from the oscillations in low order $\phi^\mathsf{P}_{\ell ,m}$ functions for $\epsilon > 0.5$.  The dotted red curve is for a narrower gap.
}
\label{fig:example}
\end{figure}

\newpage
\begin{figure}
\plotone{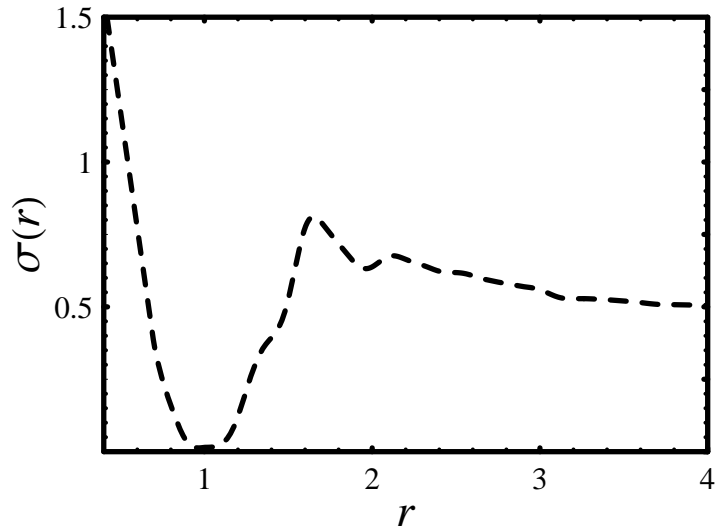}
\caption{Surface density profile for a gap as calculated in Bate et al. (2003).  This profile is used to evaluate $d \epsilon / dt$ versus $\epsilon$ in Fig. \protect\ref{fig:nocorb}.}
\label{fig:mybate}
\end{figure}

\newpage
\begin{figure}
\plotone{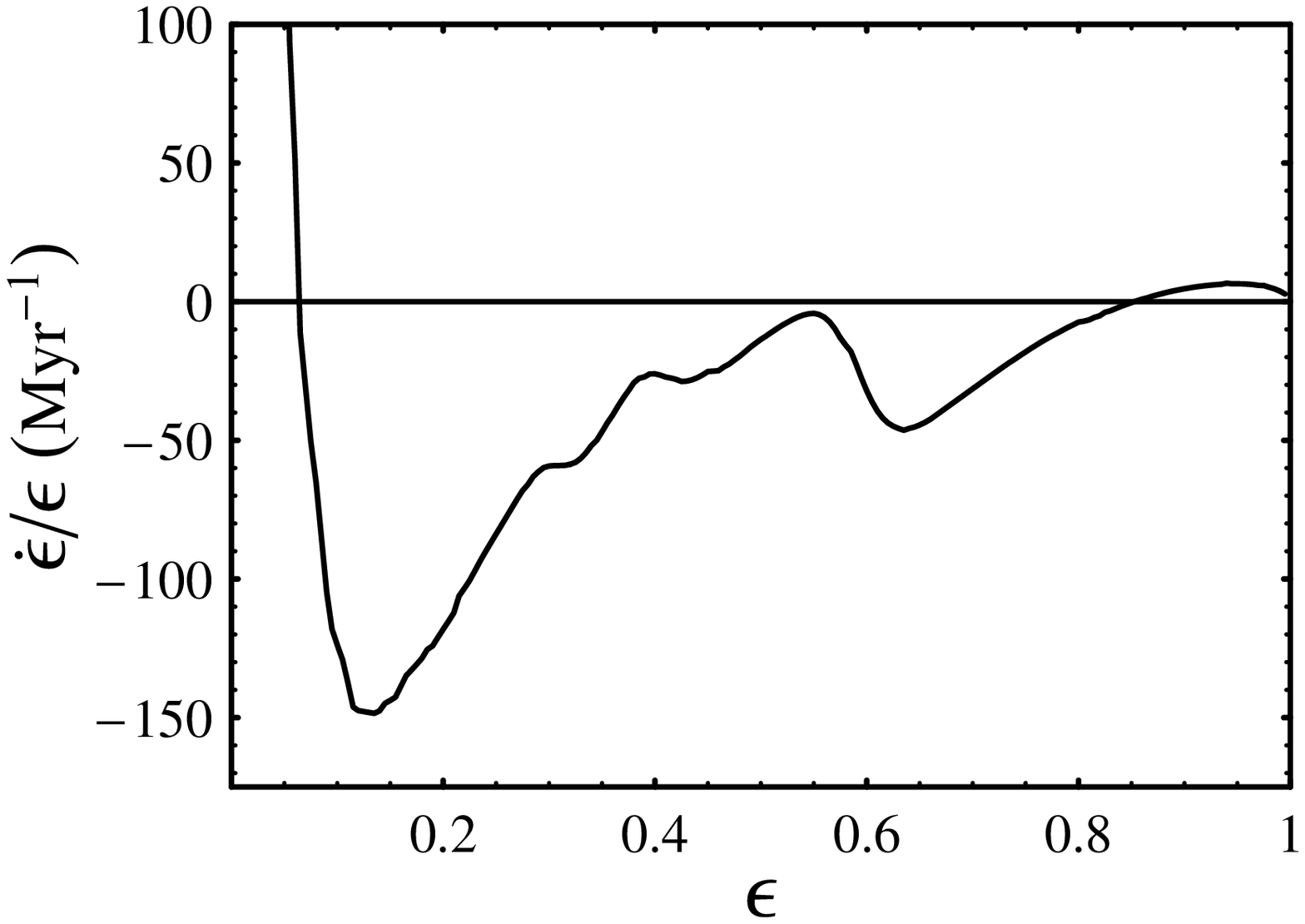}
\caption{The function $d \epsilon / dt$ versus $\epsilon$ for a 1 $M_J$ planet in a 1 AU orbit for the gap shape of Fig. \protect\ref{fig:mybate} (from Bate et al. 2003).  Here we have included non-coorbital corotation and Lindblad resonances.}
\label{fig:nocorb}
\end{figure}

\newpage
\begin{figure}
\plotone{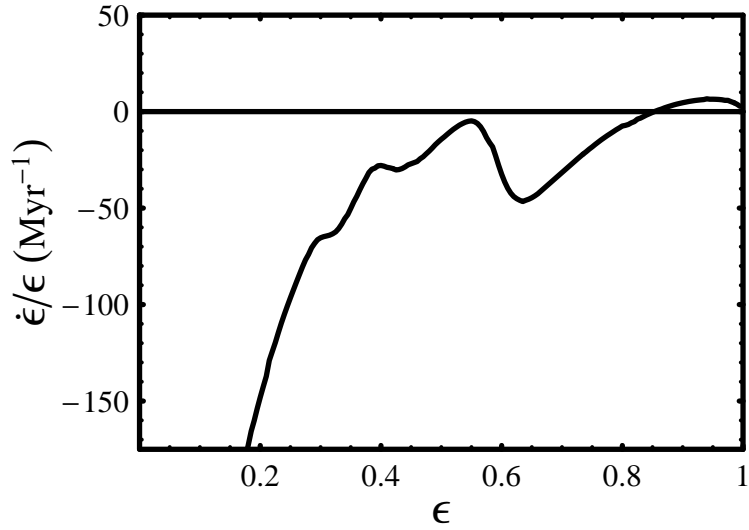}
\caption{The function $d \epsilon / dt$ versus $\epsilon$ for a 1 $M_J$ planet in a 1 AU orbit for the gap shape of Fig. \protect\ref{fig:mybate} (from Bate et al. 2003).  Here we have included coorbital corotation and Lindblad resonances as well as non-coorbital resonances.}
\label{fig:coorb}
\end{figure}

\newpage
\begin{figure}
\plotone{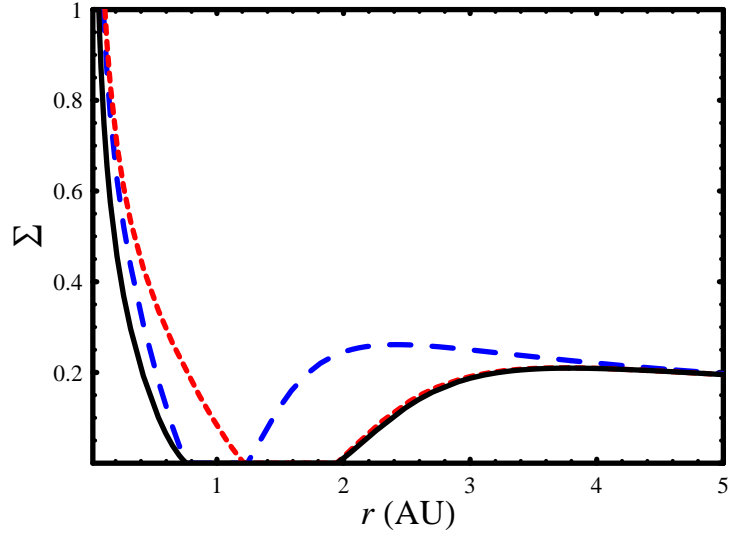}
\caption{Radial surface density profile for a disk in the presence of 1 $M_J$ planets at (dashed blue) 1 AU, (dotted red) 1.59 AU, and at both 1 AU and 1.59 AU (solid black).  This configuration places the planets in a 2-1 mean motion resonance.  The gaps produced by the two planets in the final case overlap to produce a single wide gap, in which each planet is offset from the gap center.  }
\label{fig:not}
\end{figure}

\newpage
\begin{figure}
\plotone{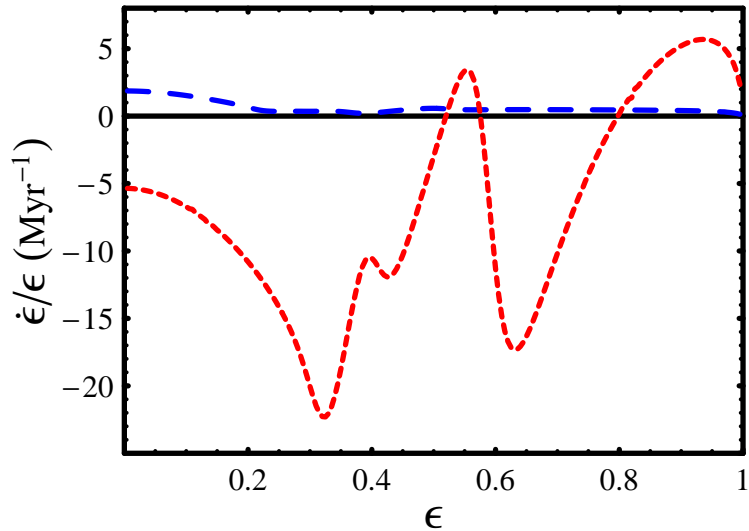}
\caption{The resulting plot of $d\epsilon /dt$ as a function of $\epsilon$ produced by the double gap shown in Fig. \protect\ref{fig:not}.   The dotted red line traces the eccentricity evolution of the outer planet, and the dashed blue line traces that of the inner planet.  The function $d\epsilon /dt$is changed substantially for the inner planet due to the lack of material at locations corresponding to the strongly contributing $m=2, \ell=1$ Lindblad and corotation resonances.}
\label{fig:2-1}
\end{figure}

\newpage
\begin{figure}
\plotone{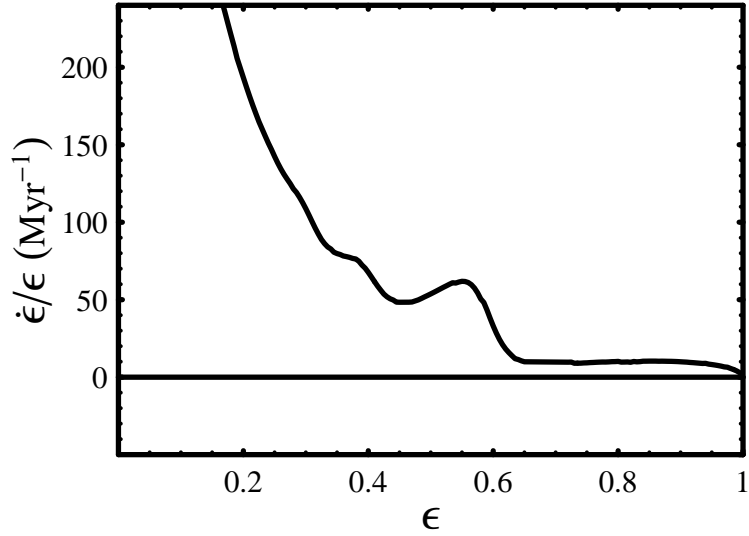}
\caption{The function $d \epsilon / dt$ versus $\epsilon$ for a 1 $M_J$ planet in a 1 AU orbit for the gap shape of Fig. \protect\ref{fig:mybate} (from Bate et al. 2003), including the effects of saturation of corotation resonances.  Here we have included only non-coorbital corotation and Lindblad resonances.  A marked difference exists between this graph and previous graphs.  This is due to the reduction of non-coorbital corotation resonant torques via corotation saturation. }
\label{fig:nocorbsat}
\end{figure}

\newpage
\begin{figure}
\plotone{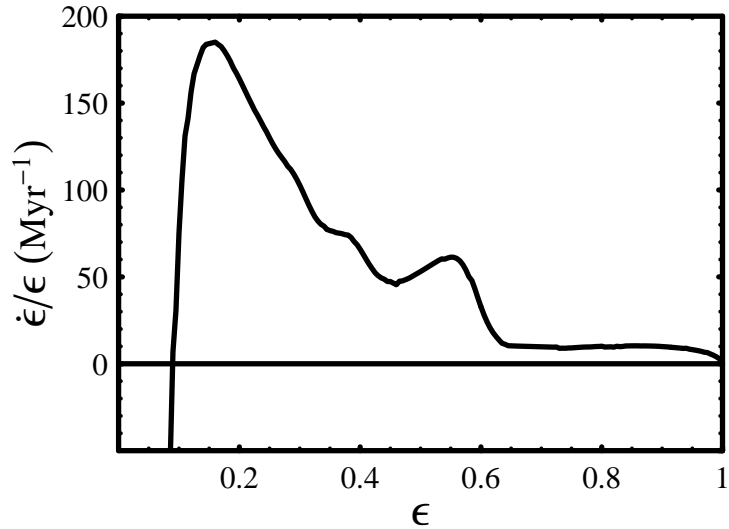}
\caption{The function $d \epsilon / dt$ versus $\epsilon$ for a 1 $M_J$ planet in a 1 AU orbit for the gap shape of Fig. \protect\ref{fig:mybate} (from Bate et al. 2003), including the effects of saturation of corotation resonances.  Here we have included coorbital corotation and Lindblad resonances as well as non-coorbital resonances, although we then assume that coorbital corotation resonances are fully saturated.}
\label{fig:scmc}
\end{figure}

\newpage
\begin{figure}
\plotone{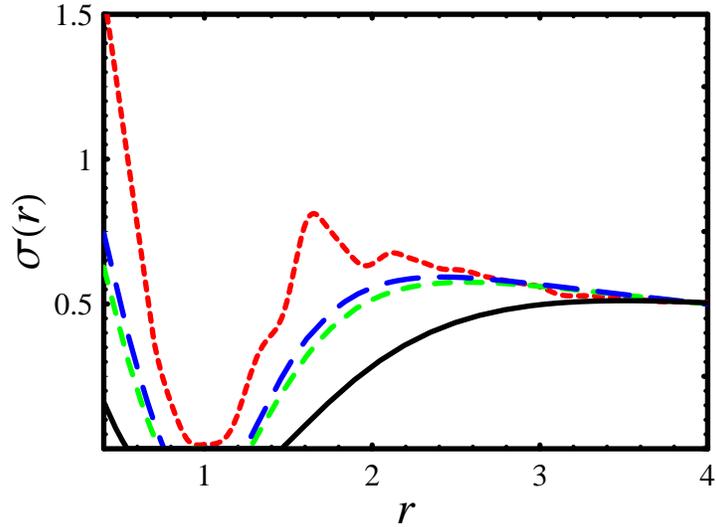}
\caption{
Surface density as a function of radius for different gap shapes corresponding to values of eccentricity evolution timescales presented in Table \protect\ref{table:edots}}
\label{fig:gaps}
\end{figure}

\newpage
\begin{figure}
\plotone{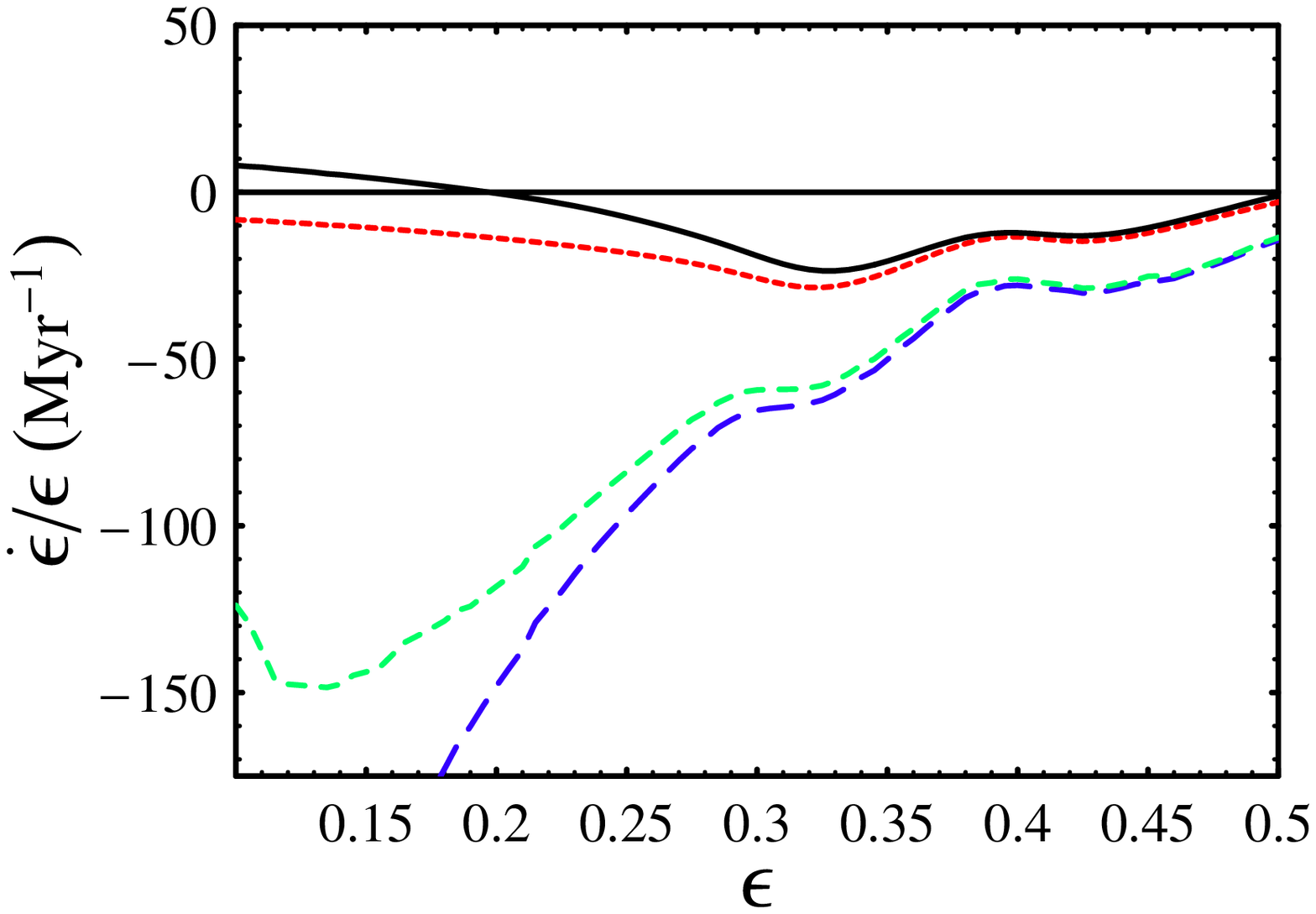}
\caption{The function $d \epsilon / dt$ versus $\epsilon$ for a 1 $M_J$ planet in a 1 AU orbit for various disk configurations.  Note that the range in eccentricity (horizontal axis) is restricted to the regime where the approximations of this paper are most applicable.  The solid black curve corresponds to a gap in the disk with sharp edges defined by $T_0 = 70$ K at $r_0 = 7$ AU, and the dotted red curve corresponds to a gap in the disk with sharp edges defined by $T_0 = 50$ K at $r_0 = 7$ AU.  The dashed green curve corresponds to the gap shape of Fig. \protect\ref{fig:mybate} (from Bate et al. 2003) with no coorbital resonances contributing, and the long-dashed blue curve to the same disk shape with coorbital resonances contributing.  In all four curves, corotation resonances are completely unsaturated.}
\label{fig:goodrange}
\end{figure}
\end{center}


\newpage
\bigskip 
\begin{thebibliography}{} 

\bibitem{al03} 
Adams, F. C., Laughlin, G., 2003.  Migration and dynamical
relaxation in crowded systems of giant planets. {Icarus}  
{163}, 290 -- 306  (AL2003). 

\bibitem{balm} 
Balmforth, N.J., Korycansky, D.G., 2001.  
Non-linear dynamics of the corotation torque. {Mon. Not. R. Astron. Soc.} {326},
833 -- 851. 

\bibitem{bate} 
Bate, M. R., Lubow, S. H., Ogilvie, G. I., Miller, K. A., 2003.
Three-dimensional calculations of high- and low-mass planets embedded
in protoplanetary discs. {Mon. Not. R. Astron. Soc.} {341},
213 -- 229. 

\bibitem{dangelo} 
D'Angelo, G., Lubow, S.H., Bate, M.R., 2006.
Evolution of Giant Planets in Eccentric Disks. {Astrophys. J.} , accepted,
astro-ph/0608355. 

\bibitem{goldreich} 
Goldreich, P., Sari, R., 2003. Eccentricity evolution for planets
in gaseous disks.  {Astrophys. J.} {585}, 1024 -- 1037. 

\bibitem{GT} 
Goldreich, P., Tremaine, S., 1980. Disk-satellite interactions. 
{Astrophys. J.} {241}, 425 -- 441. 

\bibitem{kley2004} 
Kley, W., Peitz, J., Bryden, G., 2004. Evolution of planetary systems
in resonance. {Astron. Astrophys.} {414}, 735 -- 747. 

\bibitem{laugh2004}
Laughlin, G., Steinacker, A., Adams, F., 2004. Type I Planetary Migration
with MHD Turbulence. {Astrophys. J.} {608}, 489 -- 496.

\bibitem{lee2002}
Lee, M. H., and Peale, S. J. 2002.  Dynamics and origin of the 2:1 orbital resonances
of the GJ 876 planets.  {Astrophys. J.} {567}, 596 -- 609.

\bibitem{lin}
Lin, D. N. C., Papaloizou, J. C. B., 1993.  On the tidal interaction between protostellar disks and companions.  In: Levy, E.H., Lunine, J.I. (Eds.). Protostars and Planets, vol. III. Univ. Arizona Press, Tucson, pp. 749 -- 836.

\bibitem{MB00} 
Marcy, G. W., Butler, R. P., 2000. Planets orbiting other suns. 
{Publ. Astron. Soc. Pacific} {112}, 137 -- 140.  

\bibitem{menou}
Menou, K., Goodman, J., 2004.  Low-mass planet migration in T Tauri
$\alpha$-disks.  {Astrophys. J.} {606}, 520 -- 531. 

\bibitem{ma2005}
Moorhead, A.V., Adams, F.C., 
05.  Giant planet migration through the action 
of disk torques and planet-planet scattering.  {Icarus} {178}, 517 -- 539.  

\bibitem{md} 
Murray, C. D., Dermott, S. F., 2001. {Solar System Dynamics}. 
(Cambridge: Cambridge Univ. Press). 

\bibitem{nelson}
Nelson, R. P., 2005. 
On the orbital evolution of low mass protoplanets in turbulent, magnetised disks.
{Astron. Astrophys.} {443}, 1067 -- 1085.

\bibitem{NPMK} 
Nelson, R. P., Papaloizou, J.C.B., Masset, F., Kley, W., 2000. 
The migration and growth of protoplanets in protostellar discs. 
{Mon. Not. R. Astron. Soc.} {318}, 18 -- 36.  

\bibitem{ogil} 
Ogilvie, G. I., Lubow, S. H., 2003. Saturation of the corotation
resonance in a gaseous disk.  {Astrophys. J.} {587}, 398 -- 406. 

\bibitem{papalz}
Papaloizou, J. C. B., Nelson, R. P., Masset, F., 2001.  Orbital
eccentricity growth through disc-companion tidal interaction.
{Astron. Astrophys.} {366}, 263 -- 275.

\bibitem{rasio}
Rasio, F. A., Ford, E. B., 1996. Dynamical instabilities and the 
formation of extrasolar planetary systems. {Science} {274}, 
954 -- 956. 

\bibitem{Shu} 
Shu, F. H., 1992. {Gas Dynamics}. (Mill Valley: Univ. Science Books). 

\bibitem{tj} 
Thommes, E. W., Lissauer, J. J., 2003. Resonant inclination excitation 
of migrating giant planets.  {Astrophys. J.} {597}, 566 -- 580.  

\bibitem{Trilling}
Trilling, D. E., Benz, W., Guillot, T., Lunine, J. I., Hubbard, W. B.,
Burrows, A., 1998.  Orbital evolution and migration of giant planets:
Modeling extrasolar planets.  {Astrophys. J.} {500}, 902 -- 914.

\end{thebibliography}
\end{document}